\documentclass[letterpaper,twocolumn,10pt]{article}
\usepackage{usenix-2020-09}
\usepackage{cite}
\usepackage{tikz}
\usepackage{amsmath}
\usepackage{mathtools}
\usepackage{float}
\usepackage{booktabs}
\usepackage{multirow}
\usepackage{xurl}
\usepackage{hyperref}
\usepackage{xcolor}
\usepackage{soul}
\usepackage{circledsteps}
\usepackage{makecell}
\usepackage{adjustbox}
\def\BibTeX{{\rm B\kern-.05em{\sc i\kern-.025em b}\kern-.08em
    T\kern-.1667em\lower.7ex\hbox{E}\kern-.125emX}}

\usepackage[capitalise]{cleveref}
\usepackage{booktabs}
\usepackage{subcaption}
\usepackage{placeins}
\usepackage{mdframed}
\usepackage{seqsplit}
\usepackage{listings}
\usepackage{colortbl}
\usepackage{xspace}
\usepackage{amssymb}

\newlength{\saveparindent}
\newlength{\saveparskip}
\newenvironment{newitemize}{%
\begin{list}{\mbox{}\hspace{5pt}$\bullet$\hfill}{\labelwidth=15pt%
\labelsep=4pt \leftmargin=12pt \topsep=3pt%
\setlength{\listparindent}{\saveparindent}%
\setlength{\parsep}{\saveparskip}%
\setlength{\itemsep}{3pt} }}{\end{list}}

\newcounter{ctr}

\newenvironment{newenumerate}{%
\begin{list}{{\rm \arabic{ctr})}\hfill}{\usecounter{ctr} \labelwidth=12pt%
\labelsep=3pt \leftmargin=15pt \topsep=1pt%
\setlength{\listparindent}{\saveparindent}%
\setlength{\parsep}{\saveparskip}%
\setlength{\itemsep}{1pt} }}{\end{list}}

\newcommand{\tabfontsize}{\footnotesize}

\newcommand{\codefont}[1]{\texttt{#1}}

\newcommand{\token}[1]{\lstinline[breaklines=true]{#1}}
\newcommand{\api}[1]{\texttt{#1}}

\renewcommand{\paragraph}[1]{\vspace*{6pt}\noindent\textbf{#1}\;}

\newcommand{\BCP}{BCP\xspace}
\newcommand{\BCPs}{BCPs\xspace}
\newcommand{\BCPfull}{Business Collaboration Platforms\xspace}
\newcommand{\msteams}{Microsoft Teams\xspace}

\newcommand{\revise}[1]{{#1}}

\begin{document}

\date{}

\title{\Large \bf Experimental Security Analysis of the App Model in \\
Business Collaboration Platforms$^\dagger$}


\author{
{\rm Yunang Chen$^*$}
,
{\rm Yue Gao$^*$}
,
{\rm Nick Ceccio}
,
{\rm Rahul Chatterjee}
,
{\rm Kassem Fawaz}
,
{\rm Earlence Fernandes} 
\\[3pt] University of Wisconsin--Madison
} 


\maketitle
\pagestyle{empty}

\def\thefootnote{*}
\footnotetext{Equal Contribution. $^\dagger$Accepted in USENIX Security 2022}
\def\thefootnote{\arabic{footnote}}

\begin{abstract}

Business Collaboration Platforms like Microsoft Teams and Slack enable teamwork by supporting text chatting and third-party resource integration. A user can access online file storage, make video calls, and manage a code repository, all from within the platform, thus making them a hub for sensitive communication and resources. The key enabler for these productivity features is a third-party application model. We contribute an experimental security analysis of this model and the third-party apps. Performing this analysis is challenging because commercial platforms and their apps are closed-source systems. Our analysis methodology is to systematically investigate different types of interactions possible between apps and users. We discover that the access control model in these systems violates two fundamental security principles: least privilege and complete mediation. These violations enable a malicious app to exploit the confidentiality and integrity of user messages and third-party resources connected to the platform. We construct proof-of-concept attacks that can: (1) eavesdrop on user messages without having permission to read those messages; (2) launch fake video calls; (3) automatically merge code into repositories without user approval or involvement. Finally, we provide an analysis of countermeasures that systems like Slack and Microsoft Teams can adopt today.

\end{abstract}


\section{Introduction}\label{sec:intro}

Business Collaboration Platforms (BCPs) like Slack and Microsoft Teams are indispensable collaboration and productivity tools. Beyond multi-user chat features, BCPs enhance productivity by allowing users to integrate third-party resources. For example, users can make video calls with Zoom, store files on DropBox, chat with customers, and manage code repositories, all from within the BCP. A vibrant third-party app ecosystem allows many such integrations. Thus, BCPs not only host private communications between users but also serve as a hub for all their sensitive resources from third-party systems. As such, it is vital to understand the security and privacy properties of this emerging class of distributed multi-user collaboration platforms. 

We contribute to understanding the security of BCPs by performing an experimental analysis of the third-party app model. We focus on the app model because it allows BCPs to access sensitive data from third-party systems. Although there is work on understanding the operational security issues of BCPs (e.g., web security flaws~\cite{slack-android-password,slack-feature-withdraw}), to our knowledge, no work has examined the third-party app model. We focus our work on Slack and Microsoft Teams --- two of the most widely-used BCPs with mature app ecosystems~\cite{bcp-survey}. Furthermore, these two systems share design-level commonalities and potentially with other BCPs. Thus, any security findings are potentially broadly applicable to BCP design. 

Performing the security analysis of Slack and Microsoft Teams is challenging because these systems, including their apps, are closed-source. Specifically, apps themselves are remotely-hosted web services whose endpoints are only known to the BCP. This precludes classical analysis techniques such as source code and binary analysis or API endpoint testing. As an external party, we can only interact with apps the way a human user would --- through the BCP itself. Therefore, we focus our analysis efforts on the \emph{interactions} between apps and users, such as sending messages and reacting to them. To conduct the analysis methodically, we first systematize an access control model that describes the approaches taken by Slack and Teams using a uniform vocabulary. We then explore how an attacker can violate the access control model by experimentally studying each interaction method. 

We find that the BCP app model uses a two-level access control system consisting of the OAuth protocol and a runtime policy enforcer. Abstractly, a BCP app requests OAuth tokens to interact with categories of resources. For example, an app might request an OAuth token to read chat messages. However, this token does not entirely dictate what specific messages the app can read. Thus, the user has to specify the fine-grained access control policy at runtime. Once the user installs an app and permits it to read chat messages, the user can additionally specify that the app may read messages from specific channels (e.g., the ``usenix-security-submission'' channel). Whenever an app issues an API request to the BCP server to read a chat message from a specific channel, the access control system first verifies the OAuth token and then executes a runtime policy check to verify that the app is authorized to read from that specific channel. 

By examining each interaction method between BCP apps and users, we establish that this two-level access control system does not adequately confine third-party application behavior. Concretely, we have discovered that the BCP access control system violates two standard security principles: (1) \textit{least privilege} and (2) \textit{complete mediation}\cite{saltzer1975protection}. This allows malicious apps to escalate their privilege and violate the confidentiality and integrity of private chat messages and third-party resources connected to BCPs. To demonstrate the concrete harms posed to end-users, we introduce three attack classes for BCPs along with attack prototypes:

\noindent\textbf{\textit{{(1) App-to-App Delegation Attacks (\cref{sec:app-to-app}):}}} BCPs support apps that can interact with each other for productivity reasons, independently of human involvement. To support such meaningful interactions, the BCP access control model allows apps to act on behalf of a user. We show how malicious apps can exploit this to violate the confidentiality and integrity of resources that victim apps manage. Our proof-of-concept attacks include sending arbitrary emails on a victim's behalf, merging code pull requests, and retweeting any links using the victim's account.

\noindent\textbf{\textit{(2) User-to-App \revise{Interaction Hijacking} (\cref{sec:user-to-app}):}} BCP  apps can customize how users interact with them and with workspace features. For example, an app can introduce new `slash commands' into a workspace or manipulate how URLs get unfurled. For example, one can start a Zoom video call by entering \codefont{/zoom} on the Slack UI.
We show how a second malicious app can interfere when a user attempts to interact with a benign app, a problem similar to DNS domain squatting and voice assistant skill squatting~\cite{kumar2018skill,zhang2019dangerous}. 

\noindent\textbf{\textit{(3) App-to-User Confidentiality Violations (\Cref{sec:circumventing}):}} BCP apps interact with users by participating in any approved channels or conversations, where a human user explicitly `adds' the app as a member. BCPs  implement runtime policy checks to enforce security policies in these situations.  We show how a malicious app can exploit gaps between OAuth and these runtime mechanisms to leak private messages it does not have permission to view.

\noindent\textbf{Contributions.} 
\begin{newitemize}
    \item We contribute an experimental security analysis of the app model in two widely-used BCPs --- Microsoft Teams and Slack. To guide the analysis, we derive a common access control model for these two BCPs and then experimentally examine each interaction method between apps and users. We find that the access control model violates the principle of least privilege and complete mediation.
    \item We introduce three new attack classes that leverage this fundamental shortcoming of the access control model: app-to-app delegation attacks, user-to-app \revise{interaction hijacking}, and app-to-user confidentiality violations. We constructed proof-of-concept attacks for these classes to achieve effects such as sending arbitrary emails on behalf of victims, merging code requests, launching fake video calls with loose security settings, and stealing private messages without having the appropriate permission. In certain cases, we also demonstrate how an attacker can maintain their presence even after app uninstallation.
    \item We build tools to scrape app manifest data to estimate the potential for such attacks to occur. Of the 2,460 Slack apps we analyze, we find that 1,493 (61\%) are potentially vulnerable to delegation attacks, and 563 (23\%) request the necessary permissions to carry out these attacks. Of the 1,304 \msteams~apps we analyze, we find that 427 (33\%) are vulnerable to delegation attacks. We also find that 1,266 (51\%) Slack apps use slash commands; these apps are potentially vulnerable to both the user-to-app attacks and capable of performing user-to-app attacks.
    
\end{newitemize}

Finally, we propose a set of countermeasures that BCPs like Microsoft Teams and Slack can adopt today as a temporary solution to mitigate the attacks (\cref{sec:counter}). For example, enforcing user confirmation before every app-to-app interaction and command name collision can fix most issues, but this is undoubtedly a user-hostile solution. As a result, solutions with acceptable security and usability trade-offs necessitate rethinking the app and access control model in multi-user communication platforms.

\paragraph{Ethics and Disclosure.} We conducted all experiments inside private
workspaces with the authors as the only members. We did not exercise
cross-workspace features; thus, our investigations did not influence other
workspaces. We did not distribute or submit our test malicious apps to any
BCP app directory, so our attack did not affect BCP users other than the
authors' testing accounts.
\revise{We ethically disclosed all attacks we found
to Slack and Microsoft, both of which have confirmed their existence. Due to their view of the workspace as a trusted environment, the assumptions that social engineering is a prerequisite for the attacks, and that the workspace administrator will correctly manage app installations, these attacks do not meet their definitions of a security vulnerability.}

\begin{figure*}[tb]
    \centering
    \includegraphics[width=\linewidth]{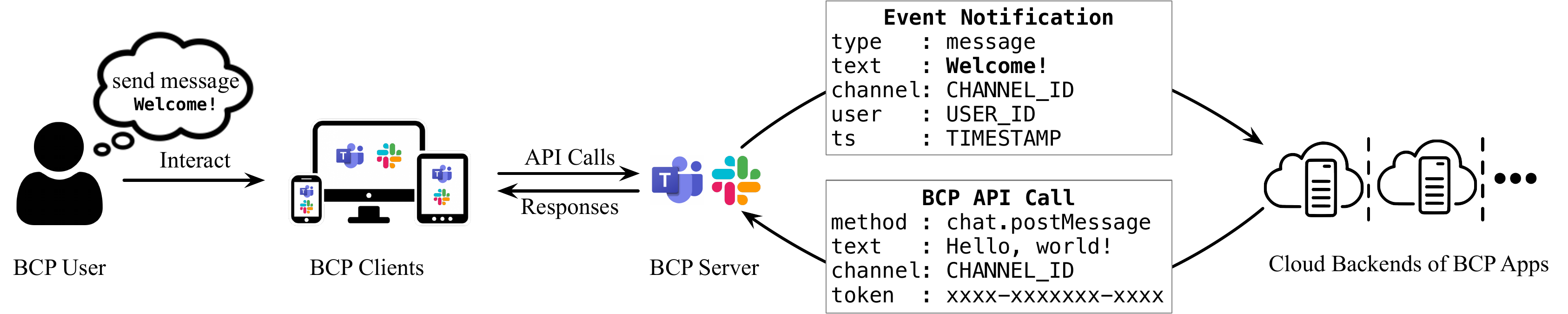}
    
    \caption{Overview of BCP's ecosystem: A \emph{BCP user} interacts with
      their \emph{BCP clients} to communicate with the \emph{BCP
        server}. BCP apps, which are maintained as separate web services by different third-party developers, communicate with
      BCP server via API calls and event
      notifications. 
      A user has to install and authorize an app before accessing its
      functionalities.}
    \label{fig:overview}
\end{figure*}

\section{\BCPfull}
\label{sec:background}

BCPs provide chatrooms that
facilitate online collaboration among a group of people, who usually belong to the same workspace, such as a project team or a research group. 
In BCPs, one can create a virtual \emph{workspace} to host all conversations for a group. 
It supports discussions among the users who joined the workspace through various conversation \emph{channels}. Users can open a new channel which can be \emph{public} --- any user can join --- or \emph{private} --- only those who are invited can join. Users can also send \emph{direct messages} to any other user or group of users in the workspace. To use a BCP, a human user interacts with their BCP client on their computer or mobile device, which then communicates with the backend servers of the BCP through various APIs. The backend server then responds to the client, updating what the user sees. We illustrate this communications framework in \Cref{fig:overview}.


In this paper, we focus on \textbf{\msteams} and \textbf{Slack}, due to their popularity and mature third-party app ecosystem. A recent survey of 900 businesses~\cite{bcp-survey} has shown that they are the two most popular BCPs\footnotemark
and are the only ones that provide a list of officially supported third-party apps.

\footnotetext{The original survey listed Skype for Business as the top spot, but it has since been  discontinued and replaced by \msteams.}



\subsection{BCP App}
\label{sec:background:app}
\label{sec:background:app.design}
Beyond basic chatting features, modern BCPs usually offer many third-party integrations, commonly known as \emph{apps}, which are cloud services providing additional productivity-enhancing functionalities in the workspace, often connecting user's data from other services (such as email or online storage) to the workspace.
These BCP apps exist on cloud servers not maintained by the BCP. These app backends communicate with the BCP servers by subscribing to event notification APIs and reacting when information about a new event is received, as depicted in \Cref{fig:overview}.
Generally, a BCP app can simultaneously act in three roles: workspace feature provider, interactive bot, and user delegate.


\paragraph{Workspace feature provider.} The app may enhance a workspace's existing features. For example, an app made by Twitter can customize the default \emph{link unfurling} feature to preview tweets linked in messages automatically. The app may also provide user-invokable actions through \emph{slash commands}. As another example, Google's Slack app \cite{gcal} shows a user's recent schedule when the user types \texttt{/gcal}.


\paragraph{Interactive bot.} The app can present itself in the workplace as a bot user and interact with other users the same way as a typical human user. The user can, for example, chat with the app's bot user directly, invite it to a channel, or share files with it. Due to these convenient features, this role has become the app's primary communication interface with its users. 

\paragraph{User delegate.} If permitted, the app may also perform actions on behalf of users. This role is particularly beneficial for enhancing productivity. For example, when users visit Dropbox's web page and wish to share files with others in their Slack workspace, they must divert their attention back and forth between Dropbox and Slack. In contrast, with the delegation ability, Dropbox enables the user to click a button without leaving the webpage and let Dropbox's Slack app \cite{dropbox} share files on their behalf. As a result, the shared files appear to have been sent directly from the user.

\subsection{Life Cycle of BCP Apps}
\label{sec:background:app.life}

\revise{\msteams{} and Slack allow any BCP user to create and distribute BCP apps without requirements, such as applying for a developer account.}
BCP apps generally go through the following stages in their life cycle: registration, publication, installation, per-user authorization, in-use, and removal. 

\paragraph{Registration.}
\label{sec:background:app.develop}
\label{sec:background:app.deploy}
To enable the various functionalities in \Cref{sec:background:app.design}, an app needs to query different web APIs or subscribe to different event notification APIs on the BCP's backend server, which in turn usually require different permissions.
%
%
The app developer must register the app in the corresponding BCP's developer portal \revise{by submitting a manifest, which specifies the app's} backend URL, required permissions, and subscribed events. 
We note that, in both \msteams~and Slack, the developer does not need to submit any of the app's codebase, as all their apps are hosted purely inside the developer's server. No client-side code is accessible by Slack, Microsoft, or the end-users.


\paragraph{Publication.}
\label{sec:background:app.publish}
After the app has been successfully registered, the developer can choose to either distribute the app\revise{'s public installation URL} through its own advertising channels or submit the app to the official app directory~\cite{slack-app-directory,teams-appsource}. 
\revise{For the second option, the app must follow submission guidelines and go through the platform's vetting procedure, which primarily involves checking if the app’s requested permissions match its claimed functionality (e.g., through a provided test account). However, as BCP apps are closed-source and their codes are not submitted for examination, it is difficult to enforce these guidelines strictly.
}



\paragraph{Installation.}
\label{sec:background:app.install}
In \msteams~and Slack, any user\footnotemark~can install an app to the workspace. During installation, a permission request page will be presented to the user, detailing what the app can do, as illustrated in \Cref{appendix:install}.
The user then either accepts all permissions or rejects all permissions. This installation is relatively invisible to other users; they are not notified when a new app is installed, and the list of installed apps is often hidden in secondary menus in the UI.

\footnotetext{Although \msteams~and Slack provide a setting for the administrators of a workspace to limit which users are allowed to install apps and which apps can be installed, the default for both BCPs is that any user can install any apps from any source. }

\paragraph{Per-User Authorization.}
\label{sec:background:app.auth}
If an app wants to act as the delegate of some users in the workspace, it may initiate a separate permission request to each user, usually by sending the request link via the app's bot user. Once the user authorizes it, the app gains permission to act on behalf of that user.  




 \paragraph{In-use and Removal.}
After the app is installed and authorized, it may additionally ask for integration with the user's account on third-party services. For example, Google's Slack app requests the user to authorize access to their Google account. BCPs do not manage the communications between BCP apps and third-party services.
If the app developer updates an app to request a different set of permissions, the user has to reinstall the app and go through the permission prompts as before. Finally, when a user uninstalls an app, \revise{it is deauthorized by the BCP}. However, there is no guarantee that the app properly disconnects itself from third-party services. 

\subsection{Security and Privacy Concerns}
\label{sec:privacy-concerns}
The widespread usage of BCPs in remote work environments implies that a lot of sensitive information passes through it. With the potential ability to access such information, BCP apps lead to security and privacy concerns. Moreover, some of the design choices that we described earlier exacerbate such concerns:
(1) \emph{all-or-nothing permissions} that disallow selective toggling of permissions; (2) \emph{imperceptible installation} that reduces the chances for users to notice what kinds of apps are installed and also prevents any workspace-wide consent mechanisms; (3) \emph{pure server-side implementation} that prevents BCPs or other entities from inspecting the app's behavior through traditional tools like static or dynamic analysis. This also allows the app to change its behavior at will.

\section{Analysis of App Permission Model in \BCP}
\label{sec:analysis}



We study the permission systems in \msteams~and Slack to identify their similarities and differences to understand the potential security design issues and systematically perform experimental security analysis.
We focus on these two \BCPs~since they are the top two most popular ones~\cite{bcp-survey} and have mature app ecosystems.
We also introduce a practical threat model and the methodology we will use to analyze the third-party apps in these two \BCPs.

\subsection{App Permission System}\label{sec:priv-model}

\begin{figure}[t]
    \centering
    \includegraphics[scale=0.56, page=1]{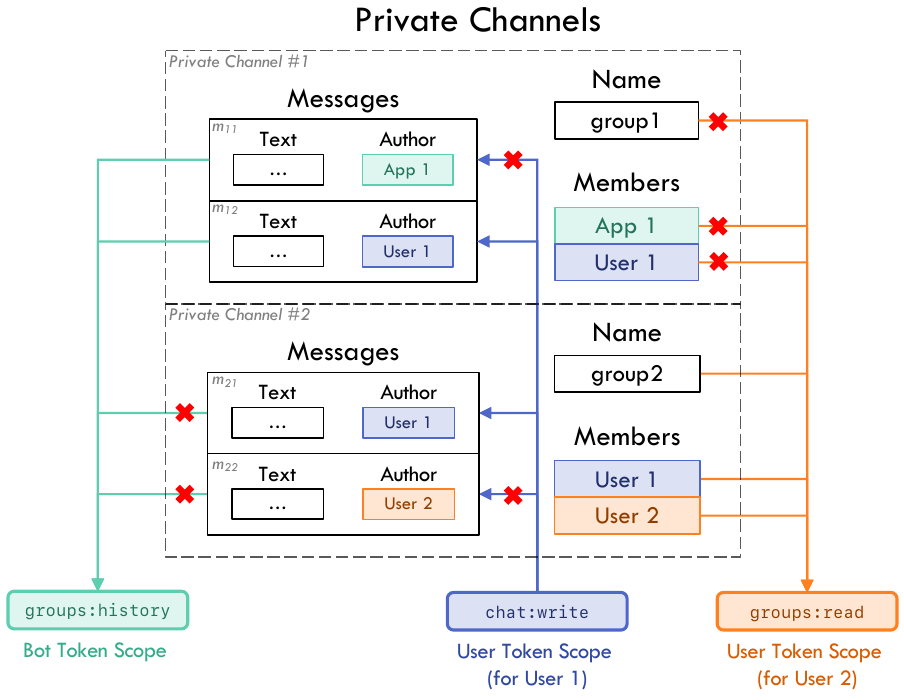}
    \caption{An example of Slack permission system. 
	We show three example scopes that App 1 may acquire.
    The arrow lines indicate that a token can be used to query all resource instances of the types allowed by the token's scope. However, Slack performs additional runtime policy checks (indicated by the red crosses) to determine which of these instances can actually be accessed. 
    }
    \label{fig:privilege_model}
\end{figure}

At a high level, \msteams~and Slack have designed their access control model based on a similar permission-based system. This permission system controls whether or not an app has access to various resources in a workspace. 
An app must first declare a set of \emph{permission scopes} it requires, with each scope representing the permission to read or write a type of resource. However, such scopes are statically defined by the \BCPs~and thus do not allow more dynamic and fine-grained access control over the specific instances under a single type of resource. To solve this problem, the \BCP~permission system includes \emph{runtime policies} that are usually user-configurable. 
For example, to read a message in a private channel, a Slack app not only needs the \texttt{groups:history} scope but also has to be added to the channel's member list by some user, as shown in \cref{fig:privilege_model}.
We now examine this two-level permission system in detail and show that it has security design issues that can violate the least privilege principle and cause privilege escalation.

\paragraph{Level 1: static permission scopes.}
An app needs to acquire several different permission scopes to perform all of its functionality. Each scope represents the permission to read or write a type of resource in a workspace, such as channel messages or shared files. 

To install the app, the user must accept all of its requested permissions; neither \BCPs~provide options to selectively toggle them.
Slack's permission scopes are implemented as standard OAuth permission scopes. Slack provides two types of scopes for its apps: \emph{bot token scope}, which allows an app to provide workspace features or act as a bot user, and  \emph{user token scope}, which allows an app to perform actions on behalf of an authorized user.
For example, the \token{chat:write} bot token scope permits the app to send messages with its bot user as the author, while the \token{chat:write} user token scope allows sending messages as the user.
\msteams follows a similar design: a set of core app capabilities that must be declared in an app's manifest is the equivalent of Slack's bot token scope, while Microsoft Graph API's OAuth permission scopes are equivalent to Slack's user token scope. The difference is that only the first type of scope is shown during the app installation; the second type can only be acquired by initiating a separate permission request to the user after installation.

These scopes are \emph{static}, in the sense that they are predefined based on how \BCPs~categorize the workspace resources, and therefore might not align with the user's desired security policies, which can vary by workspaces and evolve. To compensate for the static nature of scopes, both \BCPs~impose a second level of permission checking.


\paragraph{Level 2: runtime policy checks.}
\msteams~and Slack implement runtime policies to determine which instances in a resource type an app can access based on various conditions. Users can usually control these conditions to express their desired security policies. For example, users can have more fine-grained control of which messages in private channels an app (that has the prerequisite permission scope) can view: in Slack, they can invite the app to a specific channel, indicating that the app can view all messages inside this channel; in \msteams, they can \texttt{@}-mention the app in the messages that they wish the app to read. In this way, runtime monitors grant users some flexibility to dynamically adjust the set of resources of an app can access.


\paragraph{Security design issues.} Despite the two-level checking, we uncover two design issues in the \BCP~permission system that violate basic security principles.
\begin{enumerate}
	\item The runtime policies are ad-hoc and incomplete. 
As a result, not all user security policies can be correctly expressed. We find that not only do they differ in each \BCP, but even in the same \BCP~there are often inconsistencies between the runtime policies of similar types of resources. For example, Slack treats public channel messages and direct messages as two separate types of resources; 
however, it only imposes a policy on the former by checking whether the app is invited to the channel, but provides no mechanism to limit which user the app can send direct messages to.
The incompleteness of runtime policies leads to coarse-grained access control, violating the principle of \emph{least privilege}.

	\item The ownership or provenance of some resources is not properly tracked or enforced. This frequently happens when a user delegates an app to create resources. For example, \msteams~does not differentiate between messages sent by a real user and a delegated app. In addition, due to the multi-user multi-app nature of \BCP~workspace, the ownership of a resource can sometimes be hard to define correctly. When the ownership or provenance is absent, or the system assumes the wrong one, the principle of \emph{complete mediation} can be violated and potentially lead to privilege escalation.
\end{enumerate}
Although it is possible to build a \BCP~permission system to fix the above problems by allowing the user to specify the security policy for every instance of resources and tracking every resource's provenance, we will see in \Crefrange{sec:app-to-app}{sec:user-to-app} that such an ideal system is hard to design and often requires sacrificing usability.

\begin{figure*}[tb]
\centering\tabfontsize
\begin{tabular}{lcclp{22em}}

\toprule
  \textbf{Attack} & \textbf{Slack} & \textbf{Teams}          & \textbf{Prerequisites}    
  & \textbf{ Attack Effect Surface}             \\ 
\midrule
\textbf{Delegation}                                         & \checkmark & \checkmark              & \makecell*[{{p{19em}}}]{Permission to perform actions (primarily read \& write direct messages) on victim's behalf.}  &                   \makecell*[{{p{22em}}}]{Invoke actions in  victim's other apps to manipulate data in victim's connected third-party accounts. }    \\
-- post app removal                                        & * & \checkmark      &  \makecell*[{{p{19em}}}]{App has acquired the above permission before removal. }      &     \makecell*[{{p{22em}}}]{Incur delegation attack after the app is removed. \\
{\footnotesize *In Slack, this can only be achieved via pre-scheduled messages.}}               \\

\midrule

\textbf{Interaction hijacking}                                                    \\ 
 -- slash command                                   & \checkmark  &        &  Permission to add slash commands.                   & \makecell*[{{p{22em}}}]{Hijack any slash command in the workspace stealthily, affecting everyone using the command.     }          \\ 
 -- link unfurl                                       &  & \checkmark       & Permission to provide customized unfurling.                   & \makecell*[{{p{22em}}}]{Replace any other app's unfurled content stealthily, affecting the links sent by victim.   }            \\
\midrule
\textbf{Message extraction}                                    &                          &                              &      &   \\
-- via link unfurl                                        & \checkmark  &         &  \makecell*[{{p{19em}}}]{Permission to read \& write direct messages on victim's behalf.   }                                &  \makecell*[{{p{22em}}}]{Read messages in any private channel where victim is a member of.}  \\
-- via pin/star/reaction                                 & \checkmark  &             & \makecell*[{{p{19em}}}]{Permission to pin, star, or react to messages on victim's behalf.}                                   & \makecell*[{{p{22em}}}]{Read victim's direct messages and messages in any private channel where victim is a member of.}   \\ \bottomrule
\end{tabular}

\caption{Summary of proof-of-concept attacks and their requirements and threats. 
Per our threat model, the victim is a user who has authorized all the app's requested permissions.
} 
\label{fig:attack-summary}
\end{figure*}


\subsection{Threat Model}


Based on our analysis of the permission model above, 
we derive a threat model for \BCP apps.
We assume that the attacker has targeted a \BCP~workspace containing a number of users and already-installed apps. 
The attacker has also tricked one of the users (referred to as the victim) into installing the attacker-controlled malicious app, i.e., the victim has granted all the permission scopes requested by the malicious app.
\revise{ We believe this is a reasonable assumption, because (1) the malicious app can easily mimic a legitimate app by copying its publicly available manifest, making the two indistinguishable for the victim during installation, and (2) by default, any user in the workspace is allowed to install any app from any source.}
In our threat model, the attacker can be either an outsider or a curious user inside the workspace who wants to gain the information they cannot access.
For example, an admin can recommend everyone in the organization to install a malicious app (disguised as an innocent management app), hoping to steal chat logs from private channels they are not invited.

In addition, we assume that the \BCP's clients and its backend server are secure and do not collude with the attacker --- attacking such infrastructure is an orthogonal research direction. Therefore, the capacity of the malicious app is limited to the functionality defined by the \BCP's API. We also assume that the other apps installed in the workspace are benign and secure, which means they follow the security guidelines~\cite{slack-guidelines,teams-guidelines} and do not contain any implementation-level flaws such as exposing their tokens directly.



%

\subsection{Security Analysis Methodology}
\label{analysis:methodology}


\revise{We perform experimental security analysis on \msteams{} and Slack to study how a malicious app (defined by our threat model) can exploit the two security design issues in these two \BCPs' permission systems. Specifically, for each potential exploit, we evaluate its \emph{practicality} and \emph{prevalence}.}

\revise{To explore potential exploits}, we examine every type of interaction the malicious app can have with other entities in the workspace and check whether such interaction involves resources that have incomplete runtime policy or suffer from improper ownership tracking. If so, we explore attacks causing security-critical consequences. For each attack, we analyze how it stems from the security design issues in the permission system, how it violates the security principles, and how it jeopardizes the workspace's integrity or confidentiality guarantees expected by the user. We detail our findings in \Crefrange{sec:app-to-app}{sec:circumventing}, and summarize the prerequisites and effect surface for each attack in \cref{fig:attack-summary}.

\revise{For practicality, we build proof-of-concept malicious apps and, if applicable, target the attack on selected apps. Since most apps require a valid third-party account to function properly, running large-scale analysis is infeasible. Thus, we only select a few targeted apps that connect to sensitive resources and test them manually. We only install one targeted app at a time in our test workspace to avoid undesired interference. }

\revise{For prevalence, we analyze the app’s \emph{potential ability} to launch attacks. We collect the requested permissions of all published apps from the two \BCPs' official app category\footnotemark, and count how many apps have sufficient permissions or resources to launch each attack. It is important to note that our goal is \emph{not} to prove that some specific apps are malicious; we only examine the capabilities granted by various permission scopes and how they can be abused to perform malicious actions. This strategy allows for a sound analysis despite apps being closed-source, as the apps we find indeed have prerequisite permissions to \emph{potentially} launch attacks.}

\footnotetext{We collected 2,460 apps from the Slack~\cite{slack-app-directory} on April 7, 2021 and 1,304 apps from \msteams{}~\cite{teams-appsource} on November 17, 2021.}

\section{App-to-App Delegation Attacks}
\label{sec:app-to-app}

One of the core functionalities provided by \BCP~apps is to chat with users through their bot users interactively. However, a \BCP~app can also send and receive messages on the user's behalf and, therefore, chat with other app bot users. In this section, we present the \emph{delegation attack},  where one malicious app abuses such \emph{app-app interactions} and causes security-critical   consequences (\Crefrange{sec:app-to-app:issue}{sec:impersonation:demo}). We then show that the source of this vulnerability roots in the fundamental design issues of current \BCP~permission systems (\Cref{sec:impersonation:defense}) --- a violation of least privilege.


\subsection{App-to-App Interactions}
\label{sec:app-to-app:issue}

Both \msteams~and Slack allow their apps to present themselves in a workspace as bot users so that human users can send direct messages to these bot users to instruct them to perform certain tasks. This functionality is commonly used to let users manage their data in other online services, such as emails and file storage, without leaving the \BCP. 

At the same time, these two \BCPs~also allow apps to perform certain actions in the workspace on behalf of the user. If an app sends a message in this way, this message will appear as if the user sent it. Such delegation can be useful to enhance productivity. 
For example,  Dropbox's \BCP~app~\cite{dropbox} utilizes it to share files in channels on behalf of the user.
In Slack, this can be achieved if the app has acquired the \texttt{chat:write} user token scope in its OAuth permission request with the user; in \msteams, although none of its standard app capabilities grants permissions to delegate, one can still employ the advanced Microsoft Graph API and ask for the \texttt{Chat.ReadWrite} scope. 

By combining the above two functionalities, we can enable app-to-app interactions in \BCPs: one app that has the delegated permission to send user's messages can interact with another app's bot user. 
Such interaction can be beneficial; for example, Dokkio's Slack app~\cite{dokkio} can organize files sent by Dropbox's app into a coherent page for the workspace and tag them as shared by different users. Slack regards app-app interaction as an important feature with growing demand~\cite{goodman-wilson_2016}. 
However, allowing one app to communicate with other app's bot users has severe security implications. When the former app turns malicious, it can potentially invoke actions from the latter app, and such actions might affect data in the user's connected third-party account. We refer to attacks exploiting this vulnerability as \emph{delegation attacks}. 


We note app-app interactions can happen in other ways. Although receiving a message from the user is the most intuitive trigger event to indicate when the app should perform its actions, an app may subscribe to other triggers as well, like when  a file is shared (see \Cref{appendix:filebased}) or an emoji reaction is added. As such, apps with delegated permissions to produce these triggers can also launch potential delegation attacks.

\paragraph{Post-removal interactions. } Even after an app's removal from the workspace, it can have residual effects that cause delegation attacks. Slack provides its apps the ability to schedule a message to be sent at a future time (using the same \token{chat:write} user token scope). We find that if the app is removed before the message's scheduled time, its message will still be sent, potentially invoking actions from other apps. In \msteams, although there is no scheduling feature, this issue is more severe due to its two separate permission schemes. Upon uninstallation, only the app's standard capabilities declared in the manifest will be removed, while its delegation permissions acquired through the Graph API remain entirely intact. 
\revise{Therefore, a user \emph{cannot}, by simply removing a Teams app from the workspace, prevent the app from continuing to send messages on the user's behalf and interact with other apps, allowing the channel for delegation attacks to remain open. }

\paragraph{\revise{Current defenses.}} We note that \msteams~and Slack do have workarounds that can prevent app-to-app interactions. They allow apps to interact with users through alternative ways, such as slash commands and interactive UI windows. This prevents other apps from interfering since neither BCPs allow an app to send slash commands or click buttons in a UI. Slack in particular also tracks which messages are sent by a real user through the Slack client and which are sent by a delegated app, so that the app receiving the messages can choose whether to respond or not. However,  both of these mechanisms require the receiving app's developer to decide which actions can be triggered by other apps, but the current design of \BCP~permission system does not provide any ways for it to learn whether the delegated messages align with the user's actual intent, making it impossible to arrive at the correct decision. \revise{As we will discuss in \Cref{sec:counter}, a principled fix would trade-off functionality or usability}.

\subsection{Delegation Attack}
\label{sec:impersonation:demo}

We now focus on the delegation attack targeting both \msteams~apps and Slack apps. 
We have built a tool that crawls the information of a targeted app from the two \BCPs' official app directories and analyzes which trigger events the app is subscribing to. 
In the case of \msteams, we can also extract all message keywords that trigger the targeted app's actions.
We set up a workspace as defined per our threat model. The attacker app has acquired the appropriate delegated permission from a victim user who has also installed the targeted apps with connection to third-party services. The attacker app produces the trigger events, and we observe whether the targeted app will be tricked into performing the actions (\revise{see  \Cref{sec:implementation:app-tp-app} for more implementation details}). 
\revise{Since most apps require a valid third-party account to function properly, performing large-scale automated analysis is infeasible. Thus, in this section, we select a few apps connecting to sensitive third-party resources and manually target them, demonstrating that delegation attacks can indeed trigger security-critical or privacy-violating actions.}




\paragraph{\cstep~Send emails on victim's behalf.}
\label{sec:impersonation:demo:email}
MailClark's Slack app ~\cite{mailclark} allows sending emails directly from Slack to include non-Slack users
in a Slack conversation. 
MailClark provides a unique email address for a list of non-Slack guests in a channel configured by the user.
The email account and the recipients are only accessible to MailClark and the user.
The attacker app induces MailClark to send any emails of the attacker's choice to recipients configured by the user. Specifically, the malicious app launches this attack by sending messages to the channel as the user. During this procedure, MailClark will automatically send the attacker's message as an email to all recipients and indicate the author as the user. 

\paragraph{\cstep~Chat with victim's website visitors.}
\label{sec:impersonation:demo:livechat}
Chatlio~\cite{chatlio} is a service that lets developers add live chat functionality to their websites. It also provides an accompanying Slack app that automatically forwards any messages of the website visitors to a Slack channel and vice versa. Therefore, website owners can chat with any visitors in real-time through Slack. Unfortunately, this convenient feature makes Chatlio's app a victim of delegation attacks. Our attacker app can post messages directly into the channels used by Chatlio to chat with website visitors and thus launch further phishing attacks or harvest sensitive user info, as it now appears like a trustworthy entity to the visitors.

\paragraph{\cstep~Merge pull requests in victim's code repository.}
BitBucket's \msteams~app~\cite{bitbucket} will merge a given pull request if it receives a message starting with the keyword \texttt{merge}. It will then ask for confirmation, at which point the attacker app can reply with the text \texttt{yes} to approve the merge. The attacker app may additionally use the \texttt{list} keyword to ask BitBucket's app to display all pull requests in the victim user's connected repos or the \texttt{find} keyword to locate a specific pull request. If the repo is public, the attacker can even submit and merge its own pull request, leading to code poisoning or backdoor injection.

\paragraph{\cstep~Execute victim's automation flows.}
Microsoft Power Automate has a Teams app~\cite{powerautomate} that, upon receiving the message \texttt{Run flow [id]}, will execute the specified automation flow in the user's account. These flows can perform various actions in a wide range of services connected to Power Automate. The app also accepts messages like \texttt{List flows} and \texttt{Describe flow [id]} that can be utilized by the attacker to learn more about the user's flows and conduct more targeted attacks. 

\paragraph{\cstep~Retweet on victim's behalf.} Ziri~\cite{ziri} is a Slack app that helps users interact with tweets in a non-disruptive way. It connects to the user's Twitter account and requests permission to retweet. After that, whenever a Twitter link is shared in Slack, and the user adds a Twitter emoji reaction to that message, Ziri will automatically retweet the shared Twitter on the user's behalf. The attacker app can thus send a message containing a link to a chosen tweet (that includes harmful information) and add an emoji to the message on behalf of the user. After that, Ziri will successfully detect the tweet link and retweet it using the victim user's account. 
Such uncontrolled tweets can have detrimental effects, especially when the connected account is high profile, such as the organization's official twitter.

\paragraph{Summary.} The first four attacks rely on message events to trigger the actions in the targeted app, while the last one relies on a reaction event. We note that once the attacker and targeted apps are installed and properly authorized, the attacks do not require additional user inputs and can happen anytime, even when the user is not logged into its \BCP~client. In addition, the attacker app can delete the traces of trigger events once the attack is finished, making it even sneakier (since in both \BCPs, the permission to send messages or add emoji reactions also grants for free the permission to delete them). 

\subsection{Analysis of Root Cause and \\Potentially Prevalence}
\label{sec:impersonation:defense}

The delegation attack is possible because both \BCPs' permission systems violate the principle of \emph{least privilege}. 
Currently, the permission to send delegated messages is governed by Slack's \token{chat:write} or \msteams's \token{Chat:ReadWrite} scope; however, these two scopes allow the app to send messages to any place that the user has access to, be it a public channel, direct message with other users, or direct message with other app's bot user. In addition, neither \BCPs~provide additional runtime policies that allows the user to limit the destinations.
\revise{
Therefore, even if the user wants to install a simple app that only sends delegated messages to a small subset of other users for sharing or notification purposes, it must grant this app such overprivileged scopes that inevitable comes with the ability to launch delegation attacks. 
}

\paragraph{App's residual permissions after removal.} The reason why a removed app can still keep some residual permission differs in two \BCPs. Slack's permission system violates the \emph{principle of complete mediation} by failing to check that the proper provenance of the scheduled message, which is the removed app, should have no permissions at the time when the message is sent. Whereas in \msteams, it is the result of two separate permission systems: only the app's core capabilities are associated with Teams, while the Graph API's permissions are tied to the user's Microsoft Account (outside the permission system of Teams). Therefore, when the app is uninstalled in Teams, only the former is revoked while the latter is not affected. 
We note this issue is not Teams-specific, but also exists in other systems when permissions are managed by different trust domains~\cite{yuan2020shattered}.

\paragraph{Potential Prevalence.}
We report the number of apps capable of executing the delegation attack and that are vulnerable to the attack. 
For \msteams, we find vulnerable apps by counting apps that use bot commands capability, as these apps will accept text input from the user (or a delegated app) to perform various actions. We observe that 427 (33\%) of Teams apps use bot commands, implying that they are vulnerable to a delegation attack. However, Teams apps do not list whether they will request any delegated permission since it is acquired through a separate system. 
For Slack, we find 563 Slack apps (23\%) request
at least one `write' user scope, allowing them to interact with other apps
adversarially, while 1,493 Slack apps (61\%) request at least one `read' 
scope, implying that they are subscribing to events in the workspace and thus can be potentially affected by the attack.
We note that the measurements for Slack's vulnerable apps are the worst-case estimation. 
Since these apps are third-party web services with hidden endpoints, it is impossible to learn the app's behavior directly. 
Furthermore, most apps only perform actions after a third-party account is connected, preventing us from fully automating the evaluation of apps on a large scale.
Thus we may miscount apps that (1) have already
employed a countermeasure by blindly rejecting delegated messages, (2) subscribe the certain events but never trigger their security-critical
actions based on these events.

\section{User-to-App    Interaction    Hijacking}
\label{sec:user-to-app}

\BCPs~provide various features that serve as entry points for users to interact with apps.
Examples of these features includes `\codefont{@}'-mention, slash command, and link unfurling  (see \Cref{sec:background:app}).
In this section, we discuss how a malicious app exploits such interactions between the user and other apps in the workspace. Specifically, we find two different ways that this can happen: the malicious app can hijack other app's registered slash commands (\Cref{sec:command}), and replace another app's unfurled link content (\Cref{sec:unfurl_hijack}).
In particular, we note that both \msteams~and Slack allow apps to customize their appearance (e.g., name, icon, and description) without restriction. A malicious app can thus completely mimic the appearance of another app\footnotemark to exploit the above interactions more stealthily.
Finally, we analyze the root cause and potential prevalence of these attacks. 

\footnotetext{This may not be the case for apps published in the \BCP official catalog, as per their security guidelines. Although a Slack app can still requests \texttt{chat:write.customize} to send messages with customized appearance.}

\subsection{Slash Command Hijacking}
\label{sec:command:demo}
\label{sec:command}


In Slack's user-to-app interactions, all apps' slash commands share a single namespace, creating the potential for name collisions. A malicious app can hijack another app's commands, responding to any user that tries to launch the hijacked command in the victim app's stead. Two specific design flaws enable this attack. First, Slack only invokes the most recently installed app when multiple apps in a workspace have registered the same command. Second, both creating and renaming commands are silent and do not trigger a notification or permission prompt in Slack. As a result, one can hijack a targeted command in two ways: (1) create a new command with the same name as the targeted one; (2) rename an existing command to the targeted one.
In other words, the \texttt{commands} scope becomes over-privileged as it implicitly allows an app to take over any command within a workspace (by exploiting the name collision).  
However, Slack does not recognize this design issue as a security-critical problem\footnote{Slack acknowledged this problem in its document, but only \emph{suggests} developers to ``avoid terms that are ... likely to be duplicated,'' and not to make the command ``too complicated for users to easily remember.''}; we find no runtime policy checks of an app's permission to create or rename commands with a specific name.

\begin{figure}[tb]
    \centering
    \begin{subfigure}{0.7\linewidth}
        \centering
        \includegraphics[width=\linewidth]{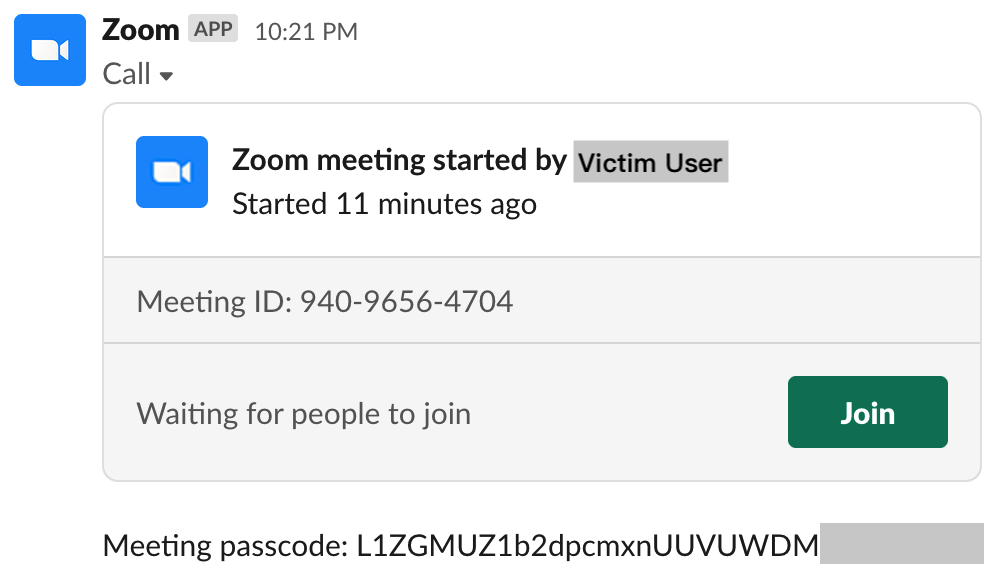}
        \caption{The official Zoom app.}
        \vspace{1em}
        \label{fig:attack:zoom:real}
    \end{subfigure}
    \begin{subfigure}{0.7\linewidth}
        \centering
        \includegraphics[width=\linewidth]{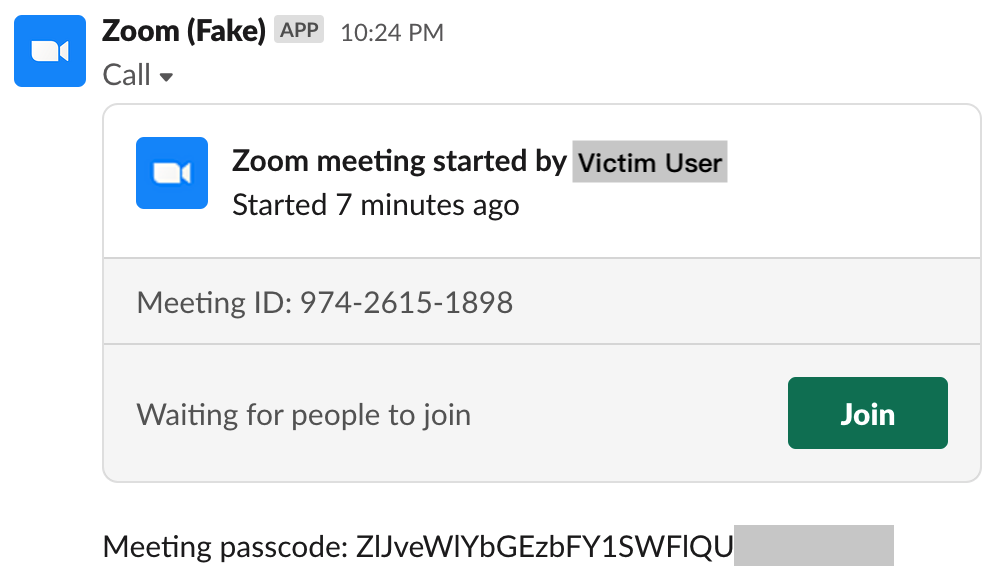}
        \caption{The spoofed Zoom meeting.}
        \label{fig:attack:zoom:fake}
      \end{subfigure}
    \caption{Zoom meetings created by official and spoofed \texttt{/zoom} commands in Slack. The spoofed Zoom meeting is secretly created by the attacker but publicly shown as started by the victim. The word ``Fake'' is added clear demonstration, it can be removed in practical attacks.}
    \label{fig:attack:zoom}
\end{figure}






We demonstrate the command hijacking attack on Zoom's Slack app~\cite{zoom}. From Zoom's app, users can invoke the command \texttt{/zoom} to start private Zoom meetings and display a Zoom call in Slack, as shown in \Cref{fig:attack:zoom:real}. If the command is invoked in a private channel, only users in this private channel will receive this private call. We create a malicious app that masquerades as the official Zoom app. At the time of installation, our malicious app requests the \texttt{commands} scope to implement a benign command called \texttt{/foo}. Once installed, we \emph{rename} this command as \texttt{/zoom} to hijack the previous official \texttt{/zoom} command. After that, the malicious app will use the attacker's Zoom account to start meetings every time a user invokes the \texttt{/zoom} command, as shown in \Cref{fig:attack:zoom:fake}. \revise{We provide more implementation details in \Cref{sec:implementation:app-tp-app}.} Attackers can also treat this vulnerability as a novel entry point for phishing attacks, as discussed in \Cref{appendix:command-phishing}.



Since \msteams~does not allow apps to register their own commands, it does not suffer from this vulnerability.

\subsection{Link Unfurling Hijacking}
\label{sec:unfurl_hijack}

\msteams allows an app to provide customized link unfurling for an authorized user. The app can register a domain in its manifest. Whenever the user posts a URL under this domain, the app can append a rich message card containing texts, images, or even interactive buttons. For example, Lucidchart's Teams app~\cite{lucidcharts} unfurls a document sharing URL to preview the document as well as a button to accept the sharing invitation. Such unfurled content can be hijacked similarly to Slack's slash command: a malicious app can register the same domain as the victim app and, if the malicious app is installed after the victim app, its unfurled content will be displayed instead of the victim app's one. Moreover, the malicious app can masquerade as the victim app to further deceive the user, as its name and icon will also be part of the unfurled content. 

While Slack also allows multiple apps to register the same domain, it chooses to display all app's unfurled contents in parallel, avoiding the issue of link unfurling hijacking.

\subsection{Analysis of Root Cause and \\
Potential Prevalence}
The command and unfurling hijacking attacks work by violating \emph{least privilege} and \emph{complete mediation}, which results from an overprivileged scope and the improper tracking of resource ownership. 
First, the corresponding scope that allows an app to use slash commands or unfurl a domain should not spontaneously grant the ability to modify the app's currently registered command names or domains; an app that performs such operation should need to be re-installed. 
Second, whenever an app registers a command or a domain, it should gain ownership of this command or domain, however, given the namespace collision, both \BCPs~fail to enforce such ownership, which thus can be easily taken over by another newly-installed app.


\paragraph{Potential Prevalence.}
In Slack, this slash command attack only exploits the \texttt{commands} scope, which is requested by 1,266 apps (51.5\%). These apps can immediately overwrite each other's commands to hijack their standard workflows. Recall, once installed, these apps can change their slash commands at any time, without requiring re-installation or notifying the users (or admins) of the workspace. We also find that many apps in the Slack App Directory already have conflicting commands: 270 apps register commands used by other apps. This implies the wide reuse of conflicting commands, and thus Slack is likely to preserve this design choice. In \msteams, the link unfurl attack relies on the 
\texttt{messageHandlers} capability, which is requested by 77 apps (5.9\%). We find that 13 of them register a domain that is also registered by other apps.

\section{App-to-User Confidentiality Violations }
\label{sec:circumventing}


We analyze the different ways in which \BCP apps interact with user messages. Our main discovery is that an attacker can leak messages from private channels without having permission to read from those channels. Concretely, we can exploit two features in Slack: (1) Link unfurling of message URLs (\Cref{sec:link-unfurl}); (2) Pinning, starring, or emoji-reacting to messages (\Cref{sec:pinned-msg}). We additionally find that the root cause behind this privilege escalation is incomplete mediation coupled with a lack of ownership tracking of resources (\Cref{sec:msg_root}). We note that in \msteams~these features are either absent or inaccessible to apps, so it does not suffer from this vulnerability.

\subsection{Message Extraction Attack via \\
Link Unfurls}
\label{sec:link-unfurl}

\BCPs~have a built-in link unfurling feature that previews the website content for any URLs contained in a chat message. We first describe how link unfurling works with message URLs and then show an attack where a malicious app \emph{without} Slack's \token{groups:history}, the permission scope that controls the read access to messages in private channel, abuses  this feature to effectively monitor all chats in any private channel joined by an authorized user.



\subsubsection{Unfurling of Message URLs}
Slack provides a public URL to every message in a workspace. This URL, if accessed, will only show the message if the login credential of a user who has access to the message is provided. 
We find that when the user sends a message $m_1$ in their own \emph{personal channel} (i.e., where users can message themselves) and $m_1$ contains a URL that links to $m_2$, where $m_2$ can be any message in any of the 
channels that the user is a member of, Slack will automatically unfurl $m_2$, adding its text content (up to 8001 characters) and author as an additional attribute to the original message $m_1$. 

\begin{figure}[t]
    \centering
    \includegraphics[scale=0.52, page=4]{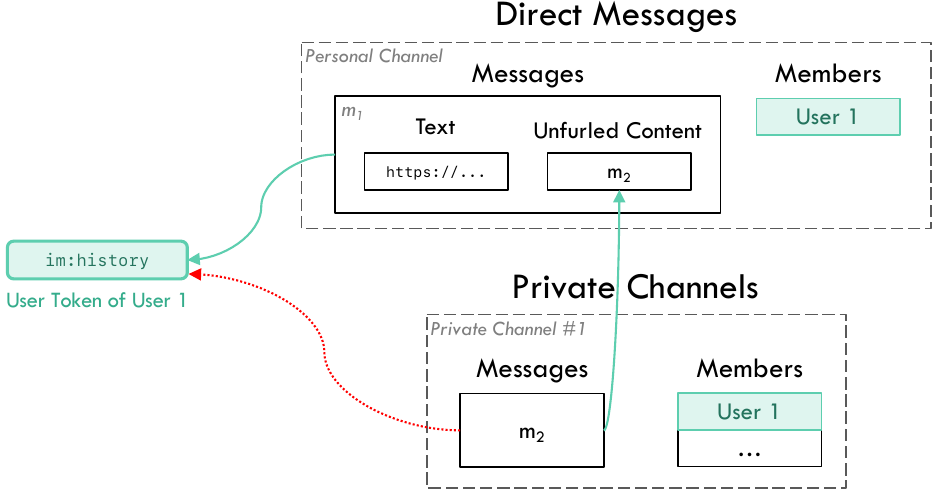}
    \caption{Privilege escalation exploiting link unfurling.
     }
    \label{fig:attack:unfurl}
\end{figure}

While this is a reasonable and useful functionality because the user's personal channel is intended for drafting messages and keeping links and files handy (as described by Slack), it leads to unwarranted access, as illustrated in  \cref{fig:attack:unfurl}. 
Slack allows an app with \token{im:history} user token scope to read the user's personal channel. 
This grants the app the ability to read $m_1$ with all its attachments.
In this case, the attachments include the unfurled content, which is $m_2$, a message from a private channel. 
Therefore, the app is implicitly permitted to read $m_2$, which is protected under the \token{groups:history} scope, and the app with only  \token{im:history} does not have access to originally.

\subsubsection{Attack Workflow}
\label{sec:link-unfurl:attack}

Now, we present a powerful attack based on the issue identified above. Through this attack, a malicious app can achieve privilege escalation --- it gains the ability to  monitor all chat messages in any private channel where the victim user is a member of,
effectively gaining the permissions provided by the \token{groups:history} user token scope but without explicitly requesting it. 




The key insight enabling this attack is that if the attacker can learn the message URL of a private channel message, it can then instruct the malicious app to  post a generated URL to the victim user's personal channel (using the \token{chat:write} scope as we described in \Cref{sec:app-to-app}), actively leaking messages from that private channel. We additionally find that Slack's message URL always follows the format:



\noindent
\begin{minipage}[c]{\linewidth}
\vspace*{6pt}
\centering
``\texttt{https://[workspace].slack.com/archives/} \\
\texttt{\textbf{[channel-ID]}/p\textbf{[message-ID]}}''
\vspace*{6pt}
\end{minipage}

\noindent 
Therefore, the attacker's job becomes learning valid combinations of channel ID and message ID. 

We have discovered several ways to obtain such combinations without resorting to \token{groups:history} and detailed them in \Cref{app:link_unfurl}. Here we describe one method that utilizes \token{groups:read}. This user token scope provides the read access to the metadata of the user's private channels, including the channel ID and the ID of the latest message in the channel.
 By constantly querying a channel's metadata, the attacker can pull every message from any private channel the victim user has joined.
We note that even if multiple messages occur between two queries, the attacker can still guess their IDs since Slack's message ID is a counter that increments for consecutive messages (see \Cref{app:link_unfurl} for details).

\paragraph{Extracting other types of messages and files.} This attack also works for other types of messages. An app's bot user can use this to view any public channel messages without the corresponding bot token scope or invitation to join that channel. 
Additionally, it can even be applied to read files shared with the user. Unlike message URL, there is no easy way to obtain a valid file URL through alternative approaches; yet, whenever a file is uploaded in a chat message, the file's public URL will also be included in that message. The attacker can then instruct Slack to unfurl the public URL to obtain a direct-downloadable link. Therefore, the attacker can access files by reading all the messages in the user's joined channels.

\subsection{Message Extraction Attack via \\
Pins, Stars, or Reactions}
\label{sec:pinned-msg}

We demonstrate another message extraction attack exploiting the incompleteness of resource ownership tracking in Slack. This time we leverage the productivity feature of pinning and starring messages (that add them to a user's saved message list) and the convenience feature of adding emoji reactions to messages.
The attack builds upon the same message ID guessing technique from the prior attack.





To pin, star, or react to a message, the app needs to present the message ID and the ID of the message's channel to the corresponding Slack API, with the \token{pins:}, \token{stars:}, or \token{reactions:write} user token scope respectively. 
However, the read counterpart of these scopes (\token{pins:}, \token{stars:}, or \token{reactions:read}) does more than permit the app to view the IDs of the pinned, starred, and reacted messages; they also allow the app to view the \emph{contents} of these messages.
Therefore, after a valid channel ID and message ID is obtained, the app with both read and write scopes can either pin, star, or react to the message, effectively allowing itself to read the given message. 
As we have seen in the prior attack, an app without permission to read a user's private channel message is still able to acquire the channel ID and message IDs of that channel's messages. Hence, a malicious app can repeatedly pin, star, or react to these messages and read through all messages in the channel.
We note that the app can also undo these operations using the corresponding write scope again to prevent the user from spotting any suspicious activity.
With this attack, the malicious app can read all the messages that the user has access to, using only these seemingly harmless operations.


\subsection{Analysis of Root Cause and \\ Potential Prevalence}
\label{sec:msg_root}

In both message extraction attacks, the malicious app obtains the ability to read any messages that the user has access to, with only some irrelevant permission scopes.   
We consider this behavior as a violation of the user's privacy expectations. When a user grants the \token{im:history} scope to an app, there is no description in the authorization prompt that suggests the app can read private channels\footnotemark. In addition, it puts the privacy of other users in these channels at risk --- the messages they posted may suddenly become accessible to an app that they never authorized. Even worse, they have no way of knowing the leakage, since all it takes is for one user to install the app, an action that is hardly perceptible to them (\Cref{sec:background:app.install}), while the app itself is never a member of the channel.

\footnotetext{Accessing private channel messages with only \token{im:history} will cause Slack API to return an \texttt{missing\_scope} error and a message saying that \token{groups:history} is needed.}

An adversarial admin can use these attacks to monitor chats in private channels they are not invited to by forcing everyone to install their malicious app that disguises itself as an innocent management app.


Such privacy violation in the first attack is a failure of not enforcing \emph{complete mediation},  which results from the improper tracking of resource provenance in Slack. Take \cref{fig:attack:unfurl} for example: when Slack finds a link to $m_2$ in $m_1$, it blindly appends the content of $m_2$ as $m_1$'s attachments, without tracking where $m_2$ originates from. As such, any entity that can read $m_1$ can also read $m_2$, whereas these two messages have different provenances and should be checked against two separate permissions. The second attack can also be mitigated if Slack tracks and checks who performed the operation. While Slack needs to allow apps to read the content of pinned, starred, or emoji-reacted messages for functionality purposes, this rule should not apply if the app trying to read the message is the one who performed the operation (since it does not make sense for an app to pin a message it does not already know).

\paragraph{Potential Prevalence.}
Out of all 1,640 apps (66.7\%) that do not request explicit scopes to read private channels (i.e., \token{groups:history}), we only counted 11 apps with the necessary permissions to extract messages via pins, stars, reactions, or link unfurls. 

%

%
%


\section{Potential Countermeasures}
\label{sec:counter}

We discuss countermeasures for the attacks we previously discussed. We note that these countermeasures are point fixes for the \BCP~permission model as it currently exists. The attack classes we've identified exist because the BCP permission model violates classic security principles. As such, even with these countermeasures, we cannot guarantee that all future issues will be prevented.
We characterize each countermeasure from three perspectives: which design issues it attempts to solve, how much it helps mitigate the attacks, and what the cost or trade-off is.

\subsection{Finer-grained Scopes}
The \BCPs~we examined define several coarse-grained scopes that manage multiple resources of different types. For example, Slack's \texttt{chat:write} user scope allows an app to send messages to \emph{any} target with the identity of the authorizing user.
The \msteams Graph API \token{Chat.ReadWrite} scope grants a \msteams~app similar permissions.
Therefore, even if the app's functionality only requires sending messages to human users, it needs to acquire one of these broad scopes, which inevitably comes with the permission to send messages to apps and thus the ability to perform impersonation attacks on other apps.
These scopes are coarse-grained as they allow an app to send messages to \emph{separate} targets (app and non-app). \BCPs~can break down these scopes into two separate scopes: one that allows sending messages to non-app targets, and another that allows messages to app targets. However, this countermeasure cannot handle the attacks exploiting scopes that do not have finer-grained concepts (such as command hijacking).

\subsection{Stricter Runtime Policy Checks}
\label{sec:counter:runtime}


Stricter runtime checks can help address the message extraction attacks found in Slack. Specifically, Slack first needs to fix its coarse-grained modeling of the message resources by decoupling the unfurled content from the message and treating it as a separate type of resource. Slack also needs to track the origin of the unfurled content, for example, whether it is a message from another channel or a file shared with the user. Then, whenever an app requests to read a message, Slack should enforce an additional dynamic condition check to examine whether the provided token has the correct privilege to access the origin of the unfurled content. If not, only the message should be returned to the app, but not the appended unfurled content.

For the attack via pins, stars, or reactions, we present two options. The first is that when an app wants to read the pinned or starred messages, Slack should send the message content only if the app has the privilege to read the original message; otherwise, only the message ID is returned.
However, this may inversely encourage malicious apps to request more privileges to maintain their original functionality.
The second is for the \BCP~to consider the entity that issued the pin, star, or react operation. For example, an app can only read the content of a pinned/starred/reacted message if the pinning/starring/reacting is done by a human user or a different app; if it is done by the requesting app itself, then the \BCP~only returns the message ID. The tracking should occur even when a user has delegated control of their account to an app. When an app performs actions on behalf of a user, those actions should still be tracked as having been taken by an app. This should not hurt any benign app's functionality because if a message is pinned, starred, or reacted on by a benign app, it is reasonable to assume that the app should already know the message's content.

However, this countermeasure does not apply to situations where it is difficult for an app or Slack to determine whether an action is malicious or user-intended. In \cref{sec:app-to-app:issue}, we demonstrated various legitimate scenarios in which users indeed want apps to perform actions on their behalf.

\subsection{Indicate Identity of Action Issuer}
\label{sec:counter:identity}
To counter delegation attacks, the victim app should be able to determine if a received event comes from a human or an impersonated user and thus choose whether to respond or not. 
Thus, \BCPs~should indicate the identity of the action issuer (i.e., whether a real or delegated user performed the action) and therefore allow for identity checks on the victim app's side.
Slack has provided this information for a few actions, such as posting messages but ignored it for other actions such as reacting to a message, which might also lead to exploits. However, as mentioned earlier, in some cases, even if the app knows the action is coming from another app, it is hard to tell whether the intent of the action is malicious or not.

\subsection{Explicit User Confirmation}
\label{sec:counter:user}
The final countermeasure is to request confirmation from users. From the perspective of victim users, all attacks stem from the fact that either victim apps or the \BCPs~automatically reacted to malicious events (in an unwanted way). Therefore, before accessing sensitive data, both the apps and the \BCP~should prompt the user for confirmation. For example, they can create a consent popup UI that involves clicking a button. Based on the current design of \msteams~and Slack, only human users can perform such actions, making it hard to forge UI actions. This will prevent both delegation and message extraction attacks.

To resolve namespace collision attacks, \BCPs should actively check for namespace collisions when apps are being installed. For example, Slack should detect when an app attempts to register a command with the same name as a command already registered in the workspace, and \msteams~should detect when an app has the same name as another app already installed in the workspace. We outline three solutions that \BCPs~may adopt. First, they can refuse to install the new app whose command would conflict with an existing one. However, this robs \BCPs~of functionality and unfairly penalizes apps installed later. Second, they can permit installation but require the user to make a selection whenever a namespace collision arises during use, but this requires the user to pay attention at all times. Third, after detecting a collision, they can provide an alias mechanism where users can change the conflicting names. In conclusion, runtime user confirmation can mitigate namespace collision attacks, but at the expense of productivity and user convenience.

\section{Related Work}

To the best of our knowledge, this is the first paper to analyze the security and privacy of third-party apps in business communication platforms. However, considerable work has been done in other types of app platforms that share varying degrees of similarities with BCPs.

\paragraph{Social networks. }
Facebook and other social network platforms allow third-party applications that offer users additional functionality and services but generally at the cost of user privacy~\cite{chaabane2014closer, robinson2014cognitive}. These apps are similar to BCP apps in terms of pure server-side implementations and all-or-nothing permission, but they are installed in a single-user home space, whereas BCP apps are in a multi-user workspace.
Symeonidis et al. show Facebook apps lead to collateral information collection~\cite{symeonidis2018collateral}, 
where they can collect not only data of the users who install them but also of their friends. This is akin to our findings of BCP apps; however, BCP apps can also actively affect other users' actions, such as through \revise{interaction hijacking}.
On the other hand, several studies propose different access control schemes for apps in social networks~\cite{anthonysamy2012collaborative, viswanath2012keeping, singh2009xbook, shehab2008beyond, cheng2013preserving, tomy2016controlling}. While these solutions aim to solve the problem of coarse-grained permissions, they usually require the social network provider to host some part of the application codes, which does not suit the current communication framework of BCP apps.




\paragraph{Voice assistants.}
Amazon Alexa, a voice assistant often built into smart home devices, allows users to install third-party apps called skills. 
 Similar to BCP apps, Alexa skills often appear in the form of chatbots; however the primary way of interacting with Alexa skills is through voice commands. Studies have shown that 
 Alexa skills can be easily squatted to enable phishing attacks~\cite{kumar2018skill, zhang2019dangerous}, similar to how Slack's commands can be hijacked. However, skill squatting relies on the inherent ambiguity of voices, whereas we exploit the namespace collisions of commands.
 In an orthogonal direction, many works try to measure the privacy practices of current Alexa skills and find that many skills do not honor their privacy policy and request overprivileged access~\cite{alhadlaq1902privacy, su2020you, guo2020skillexplorer, lentzsch2021hey}.

\paragraph{Android.}
Many studies have analyzed the security and privacy of Android apps. The closest related attacks to this work are the confused deputy and collusion attacks~\cite{davi2010privilege, marforio2011application, schlegel2011soundcomber, bugiel2012towards, lu2020demystifying}. Just as in BCPs, the app-to-app communications in Android can be used with malicious intent; however, they usually aim to achieve privilege escalation to access more user data instead of attacking users' accounts in other services. 
In addition, the problem of coarse-grained permission scopes is also found in Android, granting apps powerful capabilities that can be used to exploit various vulnerabilities~\cite{jeon2012dr}.  
Meanwhile, defenses proposed for Android apps usually require static or dynamic analysis~\cite{wong2016intellidroid, enck2014taintdroid, wei2014amandroid, gordon2015information, fratantonio2016triggerscope}, making them incompatible with BCP apps, which have no client-side codes.


\paragraph{Other OAuth-based systems.} Studies have shown that overprivileged attacks are a common issue in OAuth-based systems~\cite{ho2016smart, fernandes2016security, jia2018novel, celik2018soteria, celik2018sensitive}. In addition, despite its wide adoption, OAuth is usually poorly designed and implemented by developers~\cite{chen2014oauth, wang2015vulnerability, sun2012devil}. BCPs use coarse-grained scopes for certain operations and couple them with separate runtime policy checks that we have shown to be incomplete.


%

\section{Limitations} 
\label{sec:limitations}

\revise{For ethical reasons, we did not publish our attack apps to the Slack app directory or \msteams~app store, and thus cannot comment on their vetting processes. However, we did analyze their security guidelines~\cite{slack-guidelines, teams-guidelines} for publishing apps and found no obvious restrictions that would fundamentally prevent the attacks described in this paper. These attacks rely on abusing permissions acquired for benign purposes, causing the information-limited vetting to be ineffective.} BCPs do, however, prohibit two apps from sharing the same name, making it harder for a published app to mimic the appearance of  another app; but as we noted in \Cref{sec:user-to-app}, a Slack app can circumvent this restriction by requesting the \token{chat:write.customize} permission scope, which allows the app the send messages using customized name and icon, avoiding the need to modify the app's own name and icon declared in the manifest.

\section{Conclusions}



We performed an experimental security analysis of the app model of two popular \BCPs: Slack and \msteams. Our methodology was to study each \BCP-facilitated interaction method between apps and users. We found that these \BCPs~violate two standard security principles: least access and complete mediation. We created proof-of-concept attacks that exploit these violations to (1) impersonate users and trick victim apps into performing unwanted actions; (2) hijack commands; (3) steal messages from private channels without appropriate permissions. Our discussion of countermeasures indicates that while point fixes for these attacks can be deployed at the cost of \BCP~usability, preventing further issues requires redesigning the \BCP~app access control model.

\paragraph{Acknowledgement.}
We thank our shepherd Bruno Crispo, all anonymous reviewers and Andrei Sabelfeld for their insightful feedback.
This work was partially supported by the University of Wisconsin--Madison Office of the Vice Chancellor for Research and Graduate Education with funding from the Wisconsin Alumni Research Foundation, the DARPA GARD program under agreement number 885000, and NSF through awards:  CNS-1838733, CNS-1942014, CNS-2003129, and CNS-2144376.


\bibliographystyle{plain}
\bibliography{sample}

\FloatBarrier
\appendix

\section{Exploiting File-based Interactions}
\label{appendix:filebased}

Dokkio~\cite{dokkio} is a cloud service that provides a single place to manage a user or team's files stored in different cloud storage services, including DropBox, Google Drive, Gmail, and Slack. To manage files in Slack, it connects to the user's Slack account and request permission to read files uploaded in the workspace. Once the user shares a file in Slack, Dokkio's Slack app will automatically collect this file and provides numerous add-on services such as content organizing and cognitive services.
In this case, the user's Dokkio account is a resource that only Dokkio and the user can access. Similar to the attacks discussed above, once an app can share files on the user's behalf, it implicitly gains access to Dokkio's backend resources.

In this attack, we show that the attacker, though not authorized to access the user's Dokkio account, can add any files to the user's file management portal in Dokkio. We design a malicious app that requests the \texttt{files:write} user scope and launch the attack by uploading arbitrary files to Slack on the user's behalf. After that, Dokkio will automatically collect the shared files and add them to the user's Dokkio account.



\section{Phishing Attacks based on Command Hijacking}
\label{appendix:command-phishing}
\begin{figure}[h]
    \centering
    \includegraphics[width=\linewidth]{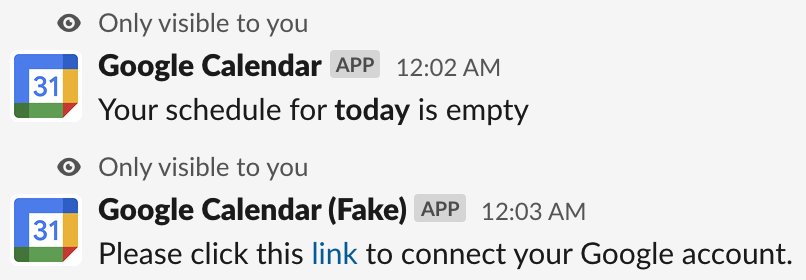}
    \caption{Demonstration of phishing attacks using the Command Hijacking attack in Slack. The two messages are sent to the user after invoking the official and hijacked \texttt{/gcal} command, respectively. The attacker can start a valid OAuth authorization process to acquire access to the user's account.}
    \label{fig:attack:gcal}
\end{figure}

Attackers can treat the design issue of command namespace collisions as a novel entry point for phishing attacks. For example, the malicious app can request the user to authorize third-party services. In \Cref{fig:attack:gcal}, we demonstrate a phishing attack by hijacking the \texttt{/gcal} command from the Google Calendar app.

\section{BCP App Installation Page}
\label{appendix:install}
\begin{figure}[h]
    \centering
    \includegraphics[width=0.75\linewidth]{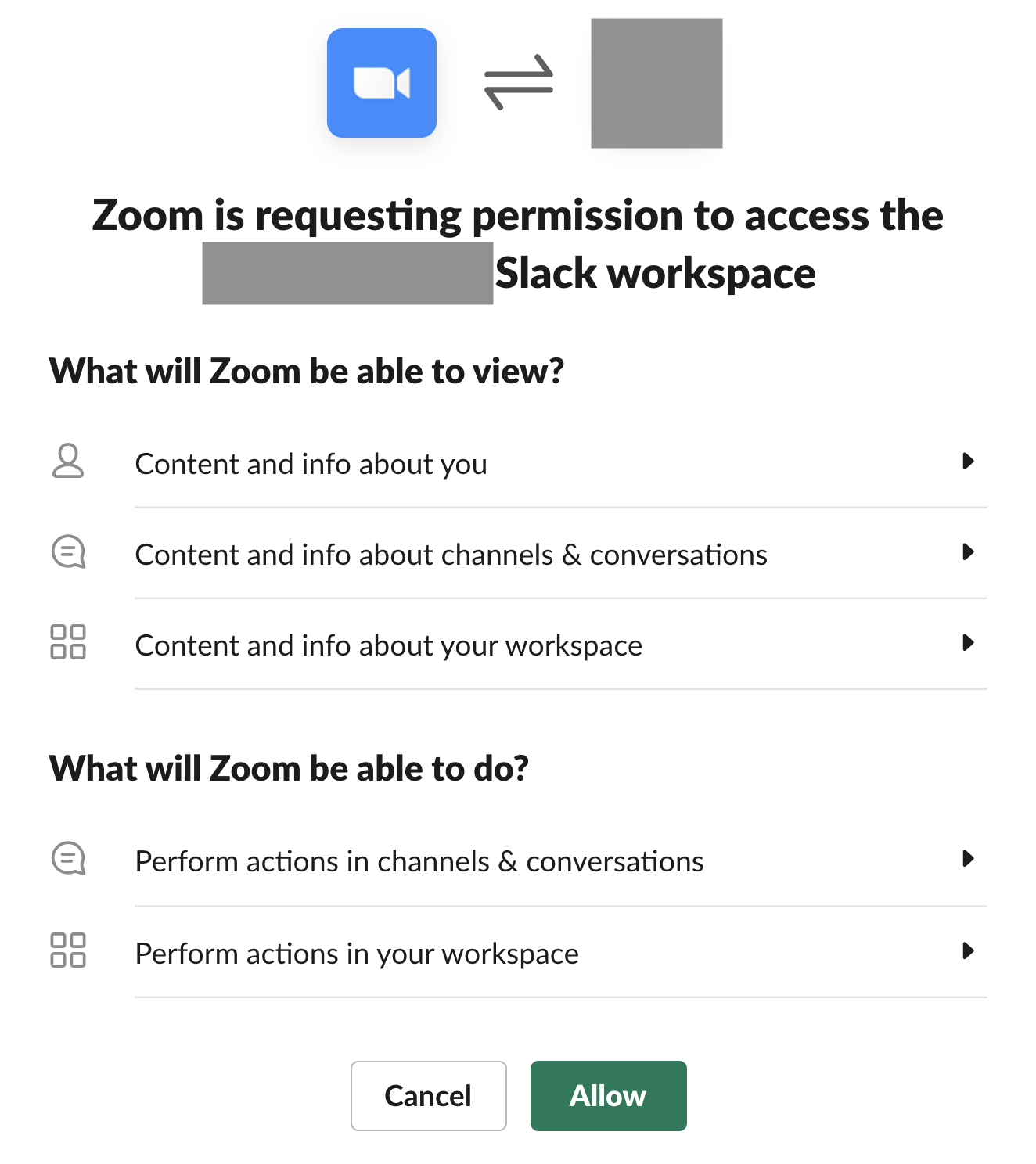}
    \caption{Installing Slack apps with bot scopes.}
    \label{fig:app-install-bot}
\end{figure}

\begin{figure}[h]
    \centering
    \includegraphics[width=0.8\linewidth]{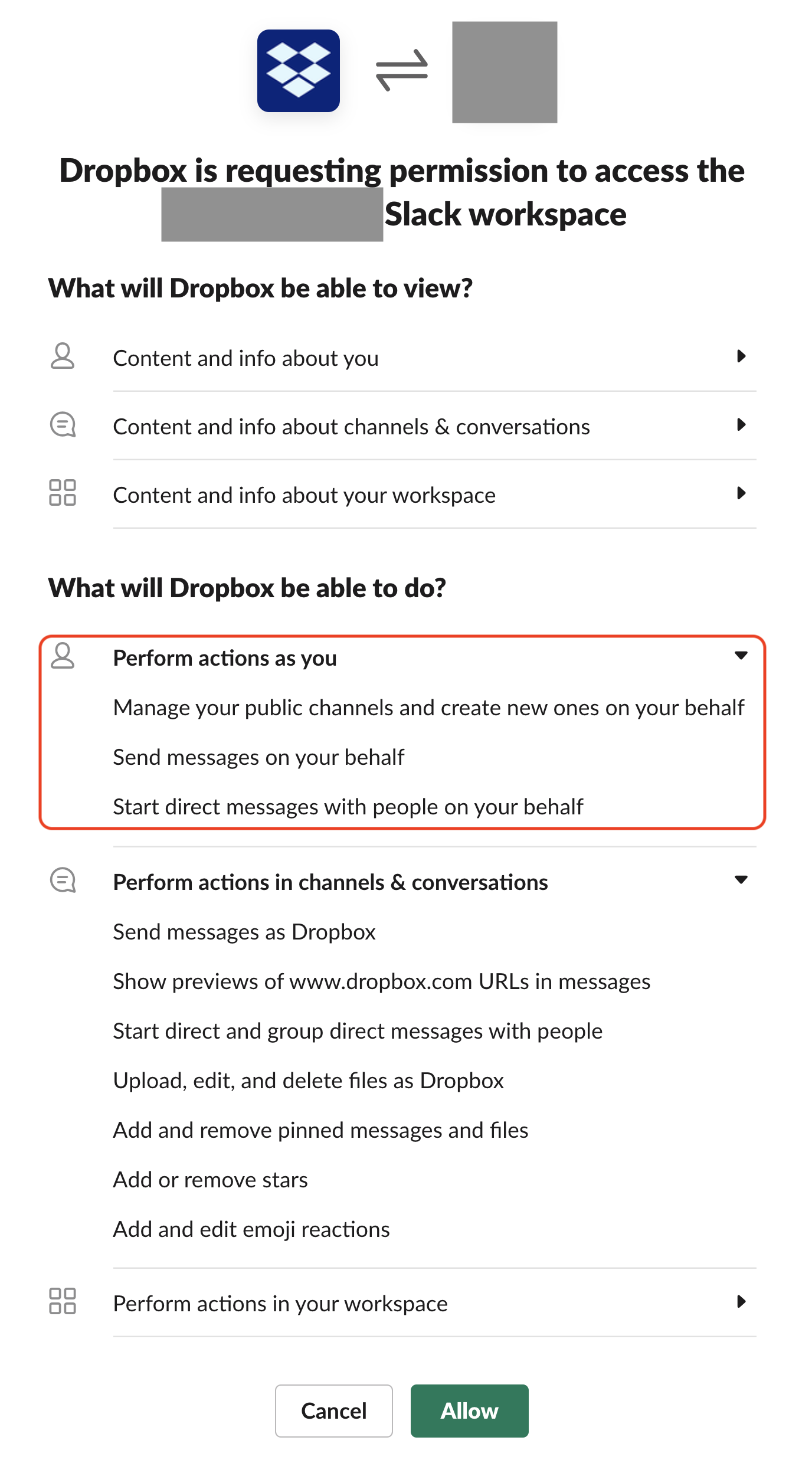}
    \caption{Installing Slack apps with user scopes.}
    \label{fig:app-install-user}
\end{figure}

\Cref{fig:app-install-bot,fig:app-install-user} show the installation of BCP apps (e.g., Slack) requesting bot scopes and user scopes. Note that in \Cref{fig:app-install-user}, the app is able to perform actions on behalf of the user, such as sending messages and direct messages.

\begin{figure}[htb]
\centering\tabfontsize
\begin{tabular}{p{16em}c}
\toprule
\textbf{User Action} & \textbf{Counter Increment} \\
\midrule
User Posting a message with text & 200 \\
App Posting a message with text & 100 \\
Posting a message with only file & 100 \\
Saving a draft (happens automatically 10 seconds after the user stops typing) & 100 \\
\bottomrule
\end{tabular}
\caption{Slack Message Counter Increment. For each consecutive message, the counter value is increased by $100x$, where $x$ starts at $0$ and gradually increases based on actions of the users in the channel.}
\label{fig:counter}
\end{figure}


\section{Implementation Details of Attacker Apps}
\label{sec:implementation}

\revise{In this section, we provide more implementation details of our attacker apps demonstrated in \Cref{sec:app-to-app,sec:user-to-app,sec:circumventing}. All apps are implemented by following the official guideline and APIs.}

\subsection{App-to-App Delegation Attacks}
\label{sec:implementation:app-tp-app}

\revise{In \Cref{sec:impersonation:demo}, we demonstrate five delegation attacks. For each attack, the attacker registers a malicious app that provides benign functionality and requests a legitimate set of permissions (detailed below). After that, the attacker either installs the malicious app to their workspace (where the attacker is a curious user) or tricks a user into installing apps in the user's workspace. Once installed and granted permission, the malicious app gets notified and starts the attack by interacting with other targeted apps in the workspace.}

\revise{The first four malicious apps request permission to send messages on behalf of the user. They launch the attack by sending specific messages that the targeted apps were designed to read and process. The last malicious app requests permission to react to messages on behalf of the user. It launches the attack by reacting with an emoji that the targeted app is designed to notice and retweet.}

\subsection{User-to-App Interaction Hijacking}
\label{sec:implementation:user-to-app}

\revise{In \Cref{sec:user-to-app}, we demonstrate the command hijacking attack on Zoom, which requires implementing a malicious app that mimics the appearance and behavior of the official Zoom app. To this end, we register an app with slash command permission but deliberately implement the command responses with Zoom APIs (of the attacker's controlled Zoom account) to mimic the official Zoom app. As BCPs permit installing apps from just a public URL, we do not have to publish the apps on official app stores. This approach avoids any accidental distribution of malicious apps to other BCP users.}

\revise{Furthermore, this attack can be extended to hijack any other apps, as long as the attacker can re-implement the proper functionalities of the targeted app. The appearance of an app is publicly available in the official app directory.}

\subsection{App-to-User Confidentiality Violations}
\label{app:link_unfurl}

We provide more details of how the attacker can obtain the channel and message IDs described in \Cref{sec:link-unfurl}.

\paragraph{Obtaining channel ID.} Each channel ID is a random string. The direct way to learn the ID of a private channel is by requesting a less alarming scope, \token{groups:read}, which provides the read access to a private channel's metadata. Alternatively, if the attacker knows the name of the channel (through side channels or guessing; per our threat model the attacker can be a curious workspace member who has some prior knowledge), it can use the \token{chat:write} scope to write a new message. It can just provide the channel name to the corresponding \api{chat.postMessage} API, which will accept this request and return the channel ID as part of the response.

\paragraph{Obtaining message ID.} The direct way to learn the message ID requires \token{groups:history}, which also grants the ability to directly read messages, avoiding the need for any attack because an app can simply misuse that permission to leak messages. However, unlike channel ID which is completely randomized, the format of a message ID follows a simple, intuitive pattern, consisting of only the current timestamp and a counter value. An example message ID is shown below: 
\[
\underbrace{\texttt{1616604187}}_{\text{\small Timestamp}}\underbrace{\texttt{0000600}}_{\text{\small Counter}}
\]
The first 10 digits represent the UNIX epoch timestamp of the message in seconds, and the last 7 digits is a counter that gets increased for each consecutive message and resets to 0 after approximately 5 days of inactivity. We conducted a series of controlled experiments and empirically found that the counter increments according to the following rules:
\begin{newenumerate}
	\item The increment between two consecutive messages is always a multiple of $100$. Although this increment is usually $200$, it may change based on the user actions listed in \cref{fig:counter}.
	\item The counters are independent across different channels, as well as user actions in different channels.
\end{newenumerate}
Due to the first rule, the attacker cannot predict the exact message ID given the previous ID, as Slack does not provide a way to learn how many drafts are saved internally. However, if the attacker is given two valid IDs separated by a small time interval, then it is straightforward to guess the valid IDs in between. 
We describe two ways of learning a valid ID. The first way is, again, to rely on the \token{groups:read} scope, since the metadata of the channel includes the ID of the latest message in the channel. The second way is to write a new message to the channel, which will cause the Slack API to return the ID of the newly posted message.

\paragraph{Attack workflow.} 

\begin{newenumerate}
	\item The attacker obtains a valid combination of channel ID and message ID using the techniques described above. We refer to the message ID as $(t_0, c_0)$. If it obtains the message ID via posting new messages, then it immediately deletes the message to hide its trace, which is also permitted by the \token{chat:write} scope.
	\item After a short time $\tau$, the attacker obtains another valid message ID $(t_0 + \tau, c_1)$. 
	\item The attacker guesses all possible message IDs, which is the cartesian product of $(t_0, t_0 + 1, ..., t_0 + \tau)$ and $(c_0 + 100, c_0 + 200, ..., c_1 - 100)$. 
	\item The attacker uses the guessed IDs to generate the message URL and posts it to the user's personal channel. The URLs of the valid IDs will get unfurled.

\end{newenumerate}

By repeating this attack over and over again for different message IDs, the attacker  can eventually pull every message from any private channel that the victim user has joined, effectively granting the malicious app the power of the \token{groups:history} scope even though this scope is never explicitly requested.
We note that the attacker should adjust the time interval $\tau$ dynamically based on the messaging frequency to aim for $c_1 - c_0 \le 500$, so that it can post all possible IDs in step 3 under Slack's rate limit (which allows unfurling of up to  5 URLs per second).

\end{document}